\documentclass{IEEEtran}
\usepackage{cite}
\usepackage{amsmath,amssymb,amsfonts}
\usepackage{graphicx}
\usepackage{textcomp,nicefrac}
\usepackage{url}
\usepackage{booktabs}
\usepackage{array}
\newcolumntype{L}[1]{>{\raggedright\arraybackslash}p{#1}}
\usepackage{tikz}
\usetikzlibrary{positioning,arrows.meta,calc,shapes.geometric,fit,backgrounds}
\def\BibTeX{{\rm B\kern-.05em{\sc i\kern-.025em b}\kern-.08em
T\kern-.1667em\lower.7ex\hbox{E}\kern-.125emX}}

\tikzset{
  blk/.style   = {draw, thick, rounded corners=1pt, align=center,
                  font=\footnotesize, minimum height=8mm, inner sep=3pt},
  mgt/.style   = {blk, fill=gray!18},
  extd/.style  = {blk, fill=gray!32},
  logic/.style = {blk, fill=white},
  hl/.style    = {blk, fill=gray!8, very thick},
  ar/.style    = {-{Stealth[length=2.2mm]}, thick},
  clkar/.style = {-{Stealth[length=2.2mm]}, very thick},
  lb/.style    = {font=\footnotesize, align=center},
  lbs/.style   = {font=\scriptsize, align=center}
}

\begin{document}
\title{TD-Link: A Daisy-Chain Optical Architecture for Integrated Data Readout and Deterministic Timing Distribution in Large-Scale Detector Systems}

\author{A.~Abba, G.~Becuzzi, M.~Bianchini, F.~Caponio, S.~Carsi, A.~Cusimano, C.~Maggio, A.~Mati, D.~Ninci, A.~Picchi, and C.~Tintori%
\thanks{Manuscript received July 1, 2026; revised October 11, 2026.}%
\thanks{A.~Abba, F.~Caponio, S.~Carsi, and A.~Cusimano are with Nuclear Instruments S.R.L., 22045 Lambrugo (CO), Italy (e-mail: abba@nuclearinstruments.eu).}%
\thanks{G.~Becuzzi, M.~Bianchini, C.~Maggio, A.~Mati, D.~Ninci, A.~Picchi, and C.~Tintori are with CAEN S.P.A., 55049 Viareggio (LU), Italy.}}

\maketitle

\begin{abstract}
TD-Link is a custom optical communication architecture that combines high-throughput data readout and sub-nanosecond timing synchronization over a single optical fiber for large-scale detector systems. The protocol adopts a multidrop daisy-chain ring topology connecting a Data Concentrator to up to sixteen FERS front-end boards per link, with up to eight independent links per concentrator. Operating at 3.125~Gb/s, TD-Link carries data, synchronization, and control traffic within the same serial stream through a token-based streaming protocol that minimizes per-hop latency and supports on-the-fly payload fragmentation. Deterministic latency is enforced at two distinct levels. On the concentrator, transmitter lane alignment is achieved by exploiting the half-full condition of the multi-gigabit transceiver (MGT) elastic buffer as a one-bit phase detector: a firmware finite-state machine (FSM) adjusts the transmit phase interpolator (PI) until the buffer write-to-read pointer difference crosses half-depth, locking every lane to the same deterministic phase condition. On each front-end board, where the synchronization pulse must additionally cross from the recovered-clock domain to the transmit domain, the residual three-to-five-cycle ambiguity of the clock-domain-crossing synchronizer is removed by a calibration-free multiphase oversampling time-to-digital converter (TDC) with 0.8~ns bins, which time-stamps the incoming synchronization edge directly on the transmit time base so that a scheduler can regenerate it at a fixed, programmable latency. A Digital Dual Mixer Time Difference (DDMTD) circuit provides the inter-concentrator phase measurement. Experimental validation with CERN PicoTDC-equipped FERS boards shows a board-to-board synchronization $\sigma$ of 24~ps for boards on the same optical ring, 27~ps across transceiver quads, and 28~ps across two cascaded concentrators, against an intrinsic front-end floor of 7~ps measured with a coaxial reference clock. The reproducibility of the chain was assessed separately, by observing the $T_0$ monitor outputs of pairs of boards on a fast oscilloscope across more than one thousand cold boots, on the same ring and on different rings, with no re-calibration of any kind: the boot-to-boot spread of the board-to-board skew remained below 20~ps, consistent with the 18.6~ps quantization of the phase-interpolator step.
\end{abstract}


\section{Introduction}
\label{sec:introduction}

\IEEEPARstart{M}{odern} large-scale detector systems for high-energy physics, nuclear physics, and medical imaging are increasingly built from many small, low-cost front-end units distributed over extended volumes. Scintillator arrays read out by silicon photomultipliers (SiPMs), wire chambers, and large solid-state detector matrices typically aggregate thousands of channels into hundreds of identical front-end boards \cite{b1,b2}. Two requirements have to be satisfied simultaneously: a sufficient aggregate data bandwidth from front-end to acquisition back-end, and a common, sub-nanosecond time reference that does not degrade as the number of boards grows.

These two requirements traditionally drive opposite design choices. Wide parallel buses or star network topologies provide ample bandwidth but require very large cable counts and dedicated timing fan-out plants. Conversely, fully serial timing distribution schemes --- such as those based on White Rabbit \cite{b3}, which extends the Precision Time Protocol \cite{b8} with synchronous Ethernet and phase measurement --- achieve sub-nanosecond accuracy but rely on a switched Ethernet infrastructure that does not naturally map onto compact, embedded detector arrays. Fixed-latency serial links over FPGA transceivers have been studied extensively in the nuclear-instrumentation community \cite{b9,b10}; the designs reported there typically bypass the transceiver elastic buffers to remove their variable contribution. Mid-size experiments and detector R\&D installations would benefit from a transport architecture that conveys data, clock, and control over the same optical fiber and that supports a daisy-chain topology where boards are physically connected one to the next.

This paper describes TD-Link, a custom 3.125~Gb/s serial protocol developed for the FERS-5200 readout platform \cite{b1}. TD-Link merges data readout and timing distribution on a single optical fiber arranged as a multidrop ring: each Data Concentrator port hosts up to sixteen FERS boards in series, and each concentrator hosts up to eight independent ports, for a maximum of 128 boards (8192~channels) per concentrator. Multiple concentrators can be cascaded coherently to instrument larger systems. Throughout this paper $T_0$ denotes the global synchronization reference of the system: a single broadcast event, encoded as a dedicated control word on the line, that simultaneously resets the time-stamp counters of every board and defines the common time origin against which all subsequent time stamps and scheduled commands are expressed. Distributing $T_0$ with a latency that is both fixed and \emph{reproducible after a power cycle} is the central design problem addressed here.

The contribution of this work is a complete, FPGA-only deterministic-latency chain that requires no per-installation calibration. Two mechanisms, acting at different points of the chain and at different time scales, are described and characterized: the elastic-buffer/phase-interpolator alignment loop on the concentrator (Section~\ref{sec:master}), and the multiphase-TDC $T_0$ regeneration on the front-end boards (Section~\ref{sec:t0regen}). The paper is organized by increasing scale. Section~\ref{sec:fers} reviews the FERS-5200 readout platform and Section~\ref{sec:protocol} the transport protocol common to both ends of the link. Section~\ref{sec:master} is devoted to the concentrator: its clock distribution, the hardware origin of the inter-lane misalignment, and the elastic-buffer alignment loop that removes it. Section~\ref{sec:slave} is devoted to the front-end board: clock recovery, word alignment, and the TDC-based $T_0$ regeneration. Section~\ref{sec:multicnc} extends the scheme to several cascaded concentrators and describes the DDMTD phase detector. Section~\ref{sec:budget} assembles the resulting per-hop latency budget, Section~\ref{sec:results} presents the experimental characterization, Section~\ref{sec:wr} compares TD-Link with White Rabbit, and Section~\ref{sec:conclusion} concludes.

\section{FERS-5200 Readout Platform}
\label{sec:fers}
FERS-5200 is a family of compact front-end readout boards developed by CAEN and Nuclear Instruments for large detector arrays, particularly those based on SiPMs, scintillators, and wire chambers. Each FERS unit provides 64 channels with on-board amplification, shaping, discrimination, charge integration, and time-to-digital conversion.

The analog front-end is built around a dedicated ASIC, and the family is defined by which one is mounted: the A5202 and A5203 boards use the Citiroc-1A, a 32-channel SiPM readout ASIC with dual-gain charge measurement and discrimination, while more recent variants of the platform adopt the Radioroc and Picoroc devices of the same family \cite{b14}, which extend the dynamic range and improve the intrinsic time resolution of the discriminator. Downstream of the ASIC, the charge information is digitized by on-board ADCs, whereas the timing information is handled by the readout device appropriate to the required resolution: a set of FPGA-based TDCs in the baseline configuration, and a CERN PicoTDC ASIC \cite{b4} on the high-resolution variant (A5203), which provides single-channel time-to-digital conversion below 10~ps and is the device used for all the measurements reported in Section~\ref{sec:results}. Acquisition logic, configuration registers, slow control of the ASICs, and the synchronization signals are managed by a Xilinx Artix-7 FPGA. Bias generation and temperature compensation of the SiPM operating point, when required, are integrated on the board.

Front-end boards connect to a Data Concentrator module --- the CAEN DT5215 --- which aggregates traffic from up to eight TD-Link ports, performs synchronization, and presents the resulting data stream to the back-end through a 1/10~Gigabit Ethernet or USB~3.0 interface (Fig.~\ref{fig:cover}). The concentrator runs the Janus data-acquisition framework \cite{b11} for configuration, monitoring, and data persistence. A small system can be deployed with a single concentrator and a handful of FERS boards; a large installation scales by adding concentrators and chaining them together coherently.

Because FERS boards are often deployed at distances of tens of meters from the back-end, and frequently inside detector volumes where shielded copper differential links are impractical, an optical link is the preferred connectivity solution: it is immune to electromagnetic interference, lightweight, and supports long distances with no signal degradation. TD-Link is the optical transport layer that meets this requirement while also acting as the timing distribution medium.

\begin{figure*}[!t]
\centering
\resizebox{7.0in}{!}{%
\begin{tikzpicture}
  \node[extd, minimum width=26mm, minimum height=20mm] (dc) at (0,0)
       {DT5215\\\textbf{Data}\\\textbf{Concentrator}\\[1mm]
        \footnotesize 8 TD-Link ports};
  \node[lbs, anchor=north] at (dc.south) {\ };
  \foreach \i/\x in {1/3.3, 2/5.3, 3/7.3} {
    \node[blk, minimum width=15mm] (a\i) at (\x,1.5)
         {FERS-5200\\\footnotesize board \i};
  }
  \node[lb] (adots) at (9.0,1.5) {$\cdots$};
  \node[blk, minimum width=15mm] (a16) at (10.8,1.5) {FERS-5200\\\footnotesize board 16};
  \draw[ar] (dc.east|-a1) -- node[above,lbs]{TD0} (a1);
  \draw[ar] (a1) -- (a2);  \draw[ar] (a2) -- (a3);
  \draw[ar] (a3) -- (adots); \draw[ar] (adots) -- (a16);
  \draw[ar] (a16.north) -- ++(0,0.75) -| node[pos=0.25,above,lbs]
       {return fiber closes the ring} ($(dc.north)+(0.6,0)$);
  \node[lbs] at (7.0,2.35) {$\times 16$ boards per ring (1024 channels)};
  \foreach \i/\x in {1/3.3, 2/5.3, 3/7.3} {
    \node[blk, minimum width=15mm] (b\i) at (\x,-1.5)
         {FERS-5200\\\footnotesize board \i};
  }
  \node[lb] (bdots) at (9.0,-1.5) {$\cdots$};
  \node[blk, minimum width=15mm] (b16) at (10.8,-1.5) {FERS-5200\\\footnotesize board 16};
  \draw[ar] (dc.east|-b1) -- node[above,lbs]{TD7} (b1);
  \draw[ar] (b1) -- (b2);  \draw[ar] (b2) -- (b3);
  \draw[ar] (b3) -- (bdots); \draw[ar] (bdots) -- (b16);
  \draw[ar] (b16.south) -- ++(0,-0.75) -| ($(dc.south)+(0.6,0)$);
  \node[lb] at (7.4,0) {$\vdots$ \quad TD1 $\ldots$ TD6 \quad $\vdots$};
  \node[extd, minimum width=22mm, minimum height=12mm] (pc) at (-4.2,0)
       {DAQ PC\\\footnotesize Janus framework};
  \draw[{Stealth[length=2.2mm]}-{Stealth[length=2.2mm]}, thick]
       (pc) -- node[above,lbs]{1/10 GbE}node[below,lbs]{or USB 3.0} (dc);
  \node[lbs, anchor=north west] at (-5.6,-2.9)
       {Single duplex fiber per ring: detector data, configuration, commands
        and the $T_0$ synchronization broadcast share the same 3.125 Gb/s stream.};
\end{tikzpicture}}
\caption{Architecture of a TD-Link readout system. The DT5215 Data Concentrator drives up to eight optical ports (TD0--TD7) at 3.125~Gb/s; each port hosts a daisy chain of up to sixteen FERS-5200 front-end boards connected one after the other along a single fiber that returns to the concentrator, closing a ring. The same fiber carries the full bidirectional traffic --- detector data, configuration, control commands, and the $T_0$ synchronization broadcast that distributes a common time reference to all boards with deterministic, sub-nanosecond latency. Aggregated data are forwarded to the DAQ PC through 1/10~Gb Ethernet or USB~3.0.}
\label{fig:cover}
\end{figure*}

\section{The TD-Link Protocol}
\label{sec:protocol}
TD-Link operates at 3.125~Gb/s using 8b/10b line coding, a rate chosen for compatibility with standard MGT primitives available in cost-effective FPGAs. With a 16-bit (two-symbol) user interface, the parallel word clock is
\begin{equation}
\label{eq:wordclk}
f_W = \frac{3.125~\mathrm{Gb/s}}{10\times 2} = 156.25~\mathrm{MHz},
\qquad T_W = 6.4~\mathrm{ns},
\end{equation}
which is the time quantum of every protocol operation described in the following. The physical medium is standard duplex small form-factor pluggable (SFP) optical transceivers and multimode fiber. Topologically, each concentrator port originates a single fiber that traverses up to sixteen FERS boards in sequence and returns to the concentrator, closing a ring. Each concentrator supports eight such rings.

The protocol multiplexes four traffic classes on the same serial bit\-stream: register configuration, control commands, synchronization commands, and detector data. Data transport is described in the following subsection; the distribution of the synchronization primitives, which is where the deterministic-latency requirement bites, is the subject of Sections~\ref{sec:master}--\ref{sec:multicnc}.

\subsection{Data Transport: Token-Based Train Protocol}
\label{sec:trainprotocol}
The multidrop ring topology imposes specific requirements on the data-transport protocol: data from each board must be aggregated on the same fiber without introducing per-hop processing delays large enough to disrupt the timing budget, and without store-and-forward buffering that would couple the link latency to the payload size. TD-Link addresses these requirements through a token-based train protocol that operates entirely in streaming mode (Fig.~\ref{fig:train}).

The concentrator periodically injects an ``empty train'' into the ring, formed by a 16-bit header (\texttt{0x8000}) and a 16-bit trailer (\texttt{0xC000}) terminated by a LAST COMMA control word. The train is forwarded by every FERS board: when the protocol FSM in the slave detects the header, it begins to retransmit the incoming stream toward the downstream node without delay, and inserts its own payload, terminated by its own cyclic redundancy check (CRC), between the upstream data and the trailer. Each board therefore appends a self-contained sub-packet with its own CRC; upstream sub-packets are forwarded verbatim and their CRC is preserved. This preserves end-to-end error detection on every individual contribution: if a payload is corrupted on any hop downstream of the originating board, the concentrator detects the error on the corresponding CRC and attributes it to the specific node. Payloads exceeding the configurable per-train fragment size are split across multiple consecutive trains.

\begin{figure*}[!t]
\centering
\resizebox{7.0in}{!}{%
\begin{tikzpicture}[x=1mm, y=1mm]
  \def\rowa{0} \def\rowb{-16} \def\rowc{-32}
  \node[lb, anchor=east] at (-2,\rowa+3) {empty train\\(concentrator)};
  \draw[thick] (0,\rowa) rectangle (16,\rowa+6);  \node[lbs] at (8,\rowa+3) {HDR\\\texttt{0x8000}};
  \draw[thick] (16,\rowa) rectangle (32,\rowa+6); \node[lbs] at (24,\rowa+3) {TRL\\\texttt{0xC000}};
  \draw[thick, fill=gray!20] (32,\rowa) rectangle (50,\rowa+6); \node[lbs] at (41,\rowa+3) {LAST\\COMMA};
  \node[lb, anchor=east] at (-2,\rowb+3) {after board 1};
  \draw[thick] (0,\rowb) rectangle (16,\rowb+6);  \node[lbs] at (8,\rowb+3) {HDR};
  \draw[thick, fill=gray!8] (16,\rowb) rectangle (44,\rowb+6); \node[lbs] at (30,\rowb+3) {payload board 1\\\footnotesize ($n$ $\times$ 32-bit words)};
  \draw[thick] (44,\rowb) rectangle (56,\rowb+6); \node[lbs] at (50,\rowb+3) {CRC$_1$};
  \draw[thick] (56,\rowb) rectangle (72,\rowb+6); \node[lbs] at (64,\rowb+3) {TRL};
  \draw[thick, fill=gray!20] (72,\rowb) rectangle (90,\rowb+6); \node[lbs] at (81,\rowb+3) {LAST\\COMMA};
  \node[lb, anchor=east] at (-2,\rowc+3) {after board 2};
  \draw[thick] (0,\rowc) rectangle (16,\rowc+6);  \node[lbs] at (8,\rowc+3) {HDR};
  \draw[thick, fill=gray!8] (16,\rowc) rectangle (44,\rowc+6); \node[lbs] at (30,\rowc+3) {payload board 1};
  \draw[thick] (44,\rowc) rectangle (56,\rowc+6); \node[lbs] at (50,\rowc+3) {CRC$_1$};
  \draw[thick, fill=gray!8] (56,\rowc) rectangle (84,\rowc+6); \node[lbs] at (70,\rowc+3) {payload board 2};
  \draw[thick] (84,\rowc) rectangle (96,\rowc+6); \node[lbs] at (90,\rowc+3) {CRC$_2$};
  \draw[thick] (96,\rowc) rectangle (112,\rowc+6); \node[lbs] at (104,\rowc+3) {TRL};
  \draw[thick, fill=gray!20] (112,\rowc) rectangle (130,\rowc+6); \node[lbs] at (121,\rowc+3) {LAST\\COMMA};
  \draw[ar] (24,\rowa-1) .. controls (24,\rowa-7) and (30,\rowb+12) .. (30,\rowb+7);
  \node[lbs, anchor=west] at (33,\rowa-6) {board 1 inserts its payload before the trailer};
  \draw[ar] (64,\rowb-1) .. controls (64,\rowb-7) and (70,\rowc+12) .. (70,\rowc+7);
  \node[lbs, anchor=west] at (73,\rowb-6) {board 2 inserts its payload, CRC$_1$ forwarded verbatim};
\end{tikzpicture}}
\caption{Token-based train protocol. The concentrator injects an empty train (header, trailer, LAST COMMA). Each board in the chain inserts its own payload and CRC immediately before the trailer while the train is streaming through, with no store-and-forward buffering: the per-hop transit time is therefore a fixed quantity, independent of the payload size. Upstream sub-packets and their CRCs are forwarded verbatim, preserving per-board end-to-end error detection.}
\label{fig:train}
\end{figure*}

This streaming, append-only behavior is what allows the per-hop transit time to be a fixed, deterministic quantity independent of payload size. The 32-bit data words generated by the FERS boards are mapped onto two consecutive 16-bit symbols on the line; control words (header, CRC, trailer, LAST COMMA) are 16-bit symbols transmitted using the K codes of the 8b/10b encoding, which keeps the framing logic simple and unambiguous. The full train returning to the concentrator is written to a per-link 64~Mword DDR4 buffer, indexed by an event pointer table, and an interrupt is raised when the buffer occupancy crosses a configurable threshold, at which point the data are forwarded to the Janus framework \cite{b11}.

At 3.125~Gb/s with 8b/10b coding the raw payload capacity of a ring is 2.5~Gb/s. The per-train overhead is four 16-bit control words (header, trailer, LAST COMMA and the inter-train gap) plus one 16-bit CRC per contributing board, i.e.\ $64 + 16 N$ bits for $N$ boards; with the default fragment size of 256 32-bit words per board the protocol efficiency exceeds 99\% for a fully populated ring, giving an effective aggregate throughput close to 2.4~Gb/s per ring and $\sim$19~Gb/s per concentrator. The measured per-hop transit latency of a streaming train is 11 word clocks, i.e.\ 70.4~ns, dominated by the 8b/10b encode/decode pipelines and by the protocol FSM; the $T_0$ fast path described in Section~\ref{sec:t0regen} is considerably shorter.

A complementary control-packet format is used in the downstream direction for register access, single-board commands, and the global $T_0$ reference. Register writes carry a 32-bit address and a 32-bit data word, split across two consecutive 16-bit link symbols, plus a CRC; register reads use the same address format and trigger a response packet from the addressed board on the next upstream train. Commands carry a 48-bit execution time stamp; each FERS board offsets the time stamp by a per-board correction and fires the command when its local counter matches. This deterministic-execution scheme allows all boards to start and stop acquisitions in lockstep regardless of their position in the chain.

\section{The Data Concentrator}
\label{sec:master}
This section describes the master end of the link: how its reference clock is generated and distributed to the eight transmitters, why the transceiver hardware makes those eight transmitters start up mutually misaligned, and how the alignment is recovered using the transmit elastic buffer as a measuring instrument.

\subsection{Clock Distribution and the Transceiver Quads}
\label{sec:masterclock}
The Data Concentrator clock distribution is summarized in Fig.~\ref{fig:masterclk}. A low-jitter, multi-output PLL (Analog Devices AD9545 \cite{b12}) is disciplined by a 10~MHz temperature-compensated crystal oscillator (TCXO); the same PLL also accepts external references, first of all from an upstream concentrator in a master/slave configuration, and additionally from upstream boards in a synchronization chain or from a GPS-disciplined receiver. In the latter case all three signals provided by the receiver are required, and they play complementary roles: the 10~MHz output disciplines the frequency of the PLL, the Pulse-Per-Second (PPS) signal marks the epoch and is the edge against which the system time base is aligned, and the NMEA sentences label that edge with the absolute date and time. Frequency without the PPS would leave the system syntonized but not synchronized; the PPS without the NMEA would give a periodic reference with no absolute label. The selected source is conditioned and used to generate two phase-coherent 156.25~MHz outputs that drive, independently, the transceiver quad PLLs (QPLLs) and the FPGA fabric.

\begin{figure*}[!t]
\centering
\resizebox{7.0in}{!}{%
\begin{tikzpicture}
  \node[extd, minimum width=20mm] (tcxo) at (0,1.5)  {10 MHz\\TCXO};
  \node[extd, minimum width=20mm] (ext)  at (0,0.3)  {external ref.\\(upstream\\concentrator)};
  \node[extd, minimum width=20mm] (gps)  at (0,-1.2) {GPS receiver\\10 MHz + PPS\\+ NMEA};
  \node[extd, minimum width=24mm, minimum height=16mm] (pll) at (3.6,0.2)
       {\textbf{AD9545}\\low-jitter PLL\\\footnotesize zero-delay mode\\\footnotesize progr.\ delay};
  \draw[ar] (tcxo) -- (pll.west|-tcxo) ;
  \draw[ar] (ext)  -- (pll.west|-ext);
  \draw[ar] (gps)  -- (pll.west|-gps);
  \node[logic, minimum width=22mm] (q0) at (9.8,2.0) {\textbf{QPLL0}\\\footnotesize 156.25 MHz $\to$ 3.125 GHz};
  \node[logic, minimum width=22mm] (q1) at (9.8,-0.3) {\textbf{QPLL1}\\\footnotesize 156.25 MHz $\to$ 3.125 GHz};
  \node[logic, minimum width=22mm] (fab) at (9.8,-2.4) {FPGA fabric\\\footnotesize TXUSRCLK 156.25 MHz};
  \coordinate (fan) at (7.6,0.2);
  \draw[very thick] (pll.east) -- node[above,lb]{$2\times$ 156.25 MHz} (fan);
  \draw[clkar] (fan) |- (q0.west);
  \draw[clkar] (fan) |- (q1.west);
  \draw[clkar] (fan) |- (fab.west);
  \node[mgt, minimum width=20mm, minimum height=13mm] (lanes03) at (14.2,2.0)
       {QUAD 0\\lanes 0--3\\\footnotesize TD0--TD3};
  \node[mgt, minimum width=20mm, minimum height=13mm] (lanes47) at (14.2,-0.3)
       {QUAD 1\\lanes 4--7\\\footnotesize TD4--TD7};
  \draw[ar] (q0) -- node[above,lbs]{$\varphi_0$ random}node[below,lbs]{at each reset} (lanes03);
  \draw[ar] (q1) -- node[above,lbs]{$\varphi_1$ random}node[below,lbs]{at each reset} (lanes47);
  \node[logic, minimum width=20mm] (ddmtd) at (14.2,-2.4) {DDMTD\\\footnotesize phase meter};
  \draw[ar] (fab) -- (ddmtd);
  \draw[ar, dashed] (fab.north) -- ++(0,0.55) -| node[pos=0.22,above,lb]
       {TXUSRCLK, common to all 8 lanes} ($(lanes47.south)+(0.5,0)$);
\end{tikzpicture}}
\caption{TD-Link master clock distribution as implemented in the DT5215 Data Concentrator. The selected source (local TCXO, external reference from an upstream concentrator, or a GPS-disciplined receiver providing 10~MHz, PPS and NMEA together) feeds the AD9545 low-jitter PLL, which generates two phase-coherent 156.25~MHz outputs: one for the transceiver QPLLs, one for the FPGA fabric. Each quad has its own QPLL, and the two QPLL outputs leave reset with independent, unpredictable phases $\varphi_0 \ne \varphi_1$; these are compensated by the per-lane transmit phase-interpolator alignment loop of Section~\ref{sec:master}, while the DDMTD provides the phase measurement used for inter-concentrator alignment (Section~\ref{sec:ddmtd}).}
\label{fig:masterclk}
\end{figure*}

The eight TD-Link transceivers of the concentrator are hosted inside the Xilinx Zynq UltraScale+ MPSoC of the DT5215, whose high-speed serial interfaces (GTH) are organized in fixed hardware groups of four called \emph{quads} \cite{b7}. Each quad is a self-contained tile that integrates four transmit/receive lanes --- parallel-in serial-out (PISO) and serial-in parallel-out (SIPO) converters, comma aligners, 8b/10b encoders, elastic buffers --- together with two LC-tank ring PLLs, the QPLLs, and one channel PLL (CPLL) per lane. The QPLLs are the only PLLs in the tile able to synthesize the multi-gigahertz line-rate clock from a sub-GHz reference (here 156.25~MHz~$\to$~3.125~GHz); they are physically located inside the quad and routed on dedicated low-skew lines to the four lanes of the same quad, but they cannot drive lanes belonging to a different quad. Two lanes hosted on different quads therefore necessarily run off two physically distinct QPLLs, each of which exits reset with an independent and random output phase. The eight TD-Link ports are consequently partitioned into two groups of four lanes (lanes~0--3 and lanes~4--7), each served by its own QPLL. This hardware-imposed partitioning is the root cause of the inter-lane phase alignment problem addressed in the rest of this section.

The consequence is that the eight optical ports of a concentrator leave reset with two unrelated transmit phases, one per quad, and that the offset between them is redrawn at every power-up. It cannot be removed by acting on the reference: an external delay placed in front of the QPLLs --- the AD9545 programmable delay, or any equivalent --- would shift the input of both QPLLs by the same amount, whereas the phase to be cancelled is generated \emph{inside} each QPLL, between its input reference and its output. A mechanism acting on a degree of freedom internal to each transceiver lane is therefore required: the transmit phase interpolator (PI).

\subsection{Principle: the Elastic Buffer as a One-Bit Phase Detector}
\label{sec:txbufferprinciple}
Before entering the implementation details, it is useful to state the principle in one sentence: \emph{the transmit elastic buffer is kept enabled, and its native half-full flag is used as a one-bit phase detector that tells the firmware on which side of the alignment point the lane currently is; the transmit PI is swept until the flag toggles, and the lane is locked on that transition.} The buffer, usually regarded as an obstacle to deterministic latency, thus becomes the measuring instrument that makes the latency deterministic, at no additional hardware cost.

\subsection{The Transmit Buffer and the Phase-Interpolator Conflict}
\label{sec:txbufferconflict}
The transmit buffer is a small FIFO between the transmit user clock domain (TXUSRCLK, the fabric clock) and the PMA serializer clock (XCLK). Its primary role is to provide safe clock-domain crossing (CDC) between two mesochronous clocks --- same frequency, unknown phase relationship after each reset. In its default configuration the FIFO adds a variable latency, because its fill level at startup depends on the random phase relationship between the two clocks; this is incompatible with the deterministic-latency requirement of the link.

A first option, and the one most commonly adopted in fixed-latency link designs \cite{b9,b10}, is to bypass the buffer entirely. In this mode the physical-coding-sublayer data path runs directly off TXUSRCLK, and clock-domain safety must be enforced by aligning the two clocks within the setup/hold window of the PISO flip-flop, using the transmit PI. The PI shifts the PMA clock in steps of about 18.6~ps referred to the 156.25~MHz clock and is therefore the natural tool to chase the random phase of the QPLL.

Unfortunately, the PI has two competing jobs when the buffer is bypassed and the lanes must also be aligned to each other, as illustrated in Fig.~\ref{fig:piconflict}. Job~1 is the CDC margin: the PI must keep TXUSRCLK and XCLK within the setup/hold window of the PISO flip-flops on every lane. Job~2 is inter-lane alignment: the PI must shift each serializer clock so that the 156.25~MHz clocks recovered at every FERS receiver are phase-aligned across all lanes. Because the QPLLs of different quads start with random phases $\varphi_0 \ne \varphi_1$, moving the PI of one lane to achieve Job~2 in general violates Job~1 on the same lane. We verified this experimentally: with the buffers bypassed and the PI steered for inter-lane alignment, the links exhibited a high rate of 8b/10b decoding and disparity errors. Buffer bypass is therefore not viable when deterministic \emph{and} lane-aligned operation is required.

\begin{figure}[!t]
\centering
\resizebox{3.45in}{!}{%
\begin{tikzpicture}[
    box/.style={draw,thick,rectangle,minimum height=9mm,minimum width=14mm,inner sep=3pt,align=center,font=\footnotesize},
    pll/.style={box,fill=gray!15,minimum width=13mm},
    piso/.style={box,fill=gray!28,minimum width=16mm},
    fab/.style={box,fill=white,minimum width=26mm},
    refsty/.style={box,fill=gray!15,minimum width=16mm},
    dat/.style={box,fill=white,minimum width=16mm},
    arrow/.style={-Stealth,thick},
    fabarrow/.style={-Stealth,thick,dashed,line width=1pt},
    >=Stealth]
\node[refsty] (ref) {156.25\,MHz\\REF CLK};
\node[pll,above right=4mm and 13mm of ref] (qpll0) {QPLL0};
\node[pll,below right=4mm and 13mm of ref] (qpll1) {QPLL1};
\node[piso,right=26mm of qpll0] (piso0) {PISO\\Lane 0};
\node[piso,right=26mm of qpll1] (piso4) {PISO\\Lane 4};
\node[dat,right=8mm of piso0] (data0) {DATA L0};
\node[dat,right=8mm of piso4] (data4) {DATA L4};
\node[fab] (txusr) at ($(piso0.south)!0.5!(piso4.north)$) {TXUSRCLK (fabric)\\$\varphi_{\rm fab}$, common to all lanes};
\draw[arrow] (ref.east) -- (qpll0.west);
\draw[arrow] (ref.east) -- (qpll1.west);
\path (ref.south) ++(0,-1.25) coordinate (refdn);
\path (txusr.west) ++(-0.7,0) coordinate (txw);
\draw[arrow] (ref.south) -- (refdn) -| (txw) -- (txusr.west);
\draw[arrow] (qpll0.east) -- node[above,font=\footnotesize]{XCLK$_0$ ($\varphi_0$)} (piso0.west);
\draw[arrow] (qpll1.east) -- node[below,font=\footnotesize]{XCLK$_1$ ($\varphi_1$)} (piso4.west);
\draw[arrow] (piso0.east) -- (data0.west);
\draw[arrow] (piso4.east) -- (data4.west);
\draw[fabarrow] (txusr.north) -- (piso0.south);
\draw[fabarrow] (txusr.south) -- (piso4.north);
\node[font=\footnotesize\bfseries,above=5mm of piso0] (ok) {CDC OK};
\draw[thick,-Stealth] (ok.south) -- (piso0.north);
\node[draw,thick,fill=gray!20,font=\footnotesize\bfseries,inner sep=3pt,align=center,below=5mm of piso4] (viol) {setup/hold\\VIOLATION};
\draw[thick,-Stealth] (viol.north) -- (piso4.south);
\end{tikzpicture}}
\caption{Dual-purpose conflict of the transmit PI in buffer-bypass mode. A single fabric clock TXUSRCLK ($\varphi_{\rm fab}$, dashed) is shared by all lanes, while each QPLL exits reset with its own random phase ($\varphi_0 \ne \varphi_1$). The PI of each lane is required to perform two jobs at once: Job~1, keep XCLK and TXUSRCLK within the setup/hold window of the PISO flip-flops; Job~2, shift each serializer clock so that the clocks recovered at the FERS receivers are phase-aligned across lanes. Because $\varphi_0 \ne \varphi_1$, satisfying Job~2 on Lane~4 in general violates Job~1 on the same lane.}
\label{fig:piconflict}
\end{figure}

\subsection{The Deterministic Buffer Solution}
\label{sec:txbuffersolution}
The solution adopted in TD-Link is to keep the elastic buffer enabled and to exploit it actively as a phase-measurement device. The transmit buffer maintains write and read pointers (WA and RA) on opposite clock domains, with Gray-coded pointer comparison to safely cross the domain boundary; its internal organization is described in the transceiver user guides \cite{b7,b13}, and Fig.~\ref{fig:txbuffer} redraws it highlighting the two elements exploited here. The native flag \texttt{txbufstatus[0]} signals whether $\mathrm{WA}-\mathrm{RA}$ is greater than or smaller than half the FIFO depth $M$:
\begin{equation}
\label{eq:halffull}
\mathrm{txbufstatus[0]} = \begin{cases} 1 & \mathrm{WA}-\mathrm{RA} \geq M/2 \\ 0 & \mathrm{WA}-\mathrm{RA} < M/2 \end{cases} .
\end{equation}
This flag is a one-bit phase detector: it tells the firmware on which side of the half-depth boundary the relative phase of the two clock domains currently sits. The transmit PI is then used to chase this flag: shifting the serializer clock advances or retards the read pointer with respect to the write pointer, and the flag toggles at the half-full crossing. Two circuits coexist in the same transceiver (Fig.~\ref{fig:txbuffer}): the \emph{actuator}, i.e.\ the transmit PI, which moves XCLK (the FIFO read clock) in steps of $\approx$1.5~ps before clock division, corresponding to 18.6~ps at the 156.25~MHz word clock, with a total range of about one word period (6.4~ns); and the \emph{sensor}, i.e.\ the fill-level comparator that produces the half-full flag used as the phase reference.

\begin{figure*}[!t]
\centering
\resizebox{7.0in}{!}{%
\begin{tikzpicture}
  \draw[very thick] (0,0) rectangle (6.4,3.2);
  \node[lb, anchor=south] at (3.2,3.3) {TX elastic buffer (depth $M$)};
  \foreach \i in {1,...,7} \draw[gray] (0,0.4*\i) -- (6.4,0.4*\i);
  \draw[fill=gray!25] (0,1.6) rectangle (6.4,3.2);
  \node[lb] at (3.2,2.4) {occupied};
  \node[lb] at (3.2,0.8) {free};
  \draw[very thick, dashed] (-0.6,1.6) -- (7.0,1.6);
  \node[lb, anchor=west] at (7.1,1.6) {$M/2$ (half-full)};
  \node[logic, minimum width=22mm] (wr) at (-3.6,2.6) {write port\\\footnotesize TXUSRCLK (fabric)};
  \draw[ar] (wr.east) -- node[above,lbs]{WA} (0,2.6);
  \node[logic, minimum width=22mm] (rd) at (10.6,0.8) {read port\\\footnotesize XCLK (PMA)};
  \draw[ar] (6.4,0.8) -- node[above,lbs]{RA} (rd.west);
  \node[hl, minimum width=30mm] (cmp) at (3.2,-2.0)
       {\textbf{SENSOR}\\fill-level comparator\\\footnotesize $\mathrm{txbufstatus[0]} = (\mathrm{WA}-\mathrm{RA} \ge M/2)$};
  \draw[ar] (3.2,0) -- (cmp.north);
  \node[hl, minimum width=30mm] (pi) at (10.6,-2.0)
       {\textbf{ACTUATOR}\\transmit phase interpolator\\\footnotesize 18.6 ps/step at 156.25 MHz};
  \draw[ar] (pi.north) -- (rd.south);
  \node[logic, minimum width=26mm] (fsm) at (6.9,-4.2) {alignment FSM\\\footnotesize sweep, then lock on $0\!\rightarrow\!1$};
  \draw[ar] (cmp.south) |- node[pos=0.28,right,lb]{1-bit phase error} (fsm.west);
  \draw[ar] (fsm.east) -| node[pos=0.72,right,lb]{PI step} (pi.south);
\end{tikzpicture}}
\caption{Deterministic TX buffer alignment loop. The \emph{sensor} is the native fill-level comparator of the elastic buffer, which emits \texttt{txbufstatus[0]} when the write-to-read pointer distance crosses half the FIFO depth; the \emph{actuator} is the transmit PI, which shifts XCLK, the FIFO read clock, in 18.6~ps steps referred to the 156.25~MHz word clock. The alignment FSM closes the loop around them: it sweeps the PI and freezes it on the flag transition, where $\mathrm{WA}-\mathrm{RA} = M/2$ exactly.}
\label{fig:txbuffer}
\end{figure*}

The alignment is performed by a small firmware FSM, summarized in Fig.~\ref{fig:fsm}. In Phase~1 (initial sweep) the FSM inspects the starting state of \texttt{txbufstatus[0]}: if the buffer is already at or above half full, the PI is decremented step-by-step until the flag drops to zero, ensuring a known starting condition below the half-full boundary; the PI is then incremented until the flag rises again from 0 to~1. In Phase~2 (lock on transition) the FSM freezes the PI on the $0\rightarrow1$ edge of \texttt{txbufstatus[0]}; this transition marks the precise half-fill point, where the write-to-read pointer difference is exactly $M/2$. The PI is held fixed afterwards, contributing no additional jitter to the link.

\begin{figure*}[!t]
\centering
\resizebox{6.4in}{!}{%
\begin{tikzpicture}[
   st/.style={draw, thick, rounded corners=3pt, align=center, font=\footnotesize,
              minimum width=32mm, minimum height=9mm, inner sep=3pt},
   dec/.style={draw, thick, diamond, aspect=2.4, align=center, font=\footnotesize,
               inner sep=1pt},
   fl/.style={-{Stealth[length=2.2mm]}, thick},
   el/.style={font=\footnotesize, align=center}]
  \node[st, fill=gray!15] (start) at (0,0) {TX reset done (start)};
  \node[dec] (d1) at (0,-2.0) {\texttt{txbufstatus[0]} $=1$ ?};
  \node[st] (dec1) at (-4.4,-4.1) {decrement PI one step};
  \node[dec] (d2) at (-4.4,-6.3) {flag $=0$ ?};
  \node[st] (inc1) at (4.4,-4.1) {increment PI one step};
  \node[dec] (d3) at (4.4,-6.3) {flag $0\!\rightarrow\!1$ ?};
  \node[st, fill=gray!25] (lock) at (4.4,-8.9) {\textbf{LOCK}: freeze PI\\$\mathrm{WA}-\mathrm{RA}=M/2$};
  \draw[fl] (start) -- (d1);
  \draw[fl] (d1) -| node[pos=0.22,above,el]{yes (at or above half)} (dec1);
  \draw[fl] (d1) -| node[pos=0.22,above,el]{no (below half)} (inc1);
  \draw[fl] (dec1) -- (d2);
  \draw[fl] (inc1) -- (d3);
  \draw[fl] (d2.west) -- (-7.6,-6.3) -- node[left,el]{no} (-7.6,-4.1) -- (dec1.west);
  \draw[fl] (d3.east) -- (7.6,-6.3) -- node[right,el]{no} (7.6,-4.1) -- (inc1.east);
  \draw[fl] (d2.south) -- (-4.4,-8.9) -- node[below,el]{yes: now below half, start sweeping up} (0.9,-8.9) -- (0.9,-4.1) -- (inc1.west);
  \draw[fl] (d3.south) -- node[right,el]{yes} (lock.north);
\end{tikzpicture}}
\caption{FSM of the deterministic TX buffer alignment. Phase~1 performs an initial sweep of the PI, starting by decrementing if the buffer is already half-full at startup, and then incrementing until the flag transitions from 0 to 1; Phase~2 locks the PI on this transition. The result is a deterministic write-to-read pointer relationship for every lane, with a residual inter-lane skew of a few PI steps.}
\label{fig:fsm}
\end{figure*}

\subsection{Why the Lock Point Is Deterministic and Does Not Drift}
\label{sec:txbufferwhy}
Two questions naturally arise about this scheme: how the half-full condition relates to the 156.25~MHz word clock, and what happens if the pointer difference moves away from $M/2$ during operation.

The half-full condition is \emph{not} configured in word-clock units, and no such configuration is needed: the flag is a structural property of the FIFO, which compares the Gray-coded pointers against the fixed threshold $M/2$. The alignment is achieved solely by PI shifts. What the PI controls is the \emph{phase} of the read clock; the pointer distance is the consequence. Sweeping the PI over one word period moves the read pointer by exactly one position relative to the write pointer, so the flag toggles once per sweep and the $0\rightarrow1$ transition is a unique, well-defined point within the range. The quantization of the lock point is therefore set by the PI step (18.6~ps at the word clock), not by the 6.4~ns word period: the residual uncertainty of the alignment is sub-nanosecond by construction, and the measured lane-to-lane spread after lock is of the order of one PI step, i.e.\ below 20~ps.

The pointer difference cannot drift away from $M/2$ during operation, because the write and read clocks are strictly mesochronous: both derive from the same AD9545 output through the same QPLL, so their frequencies are identical by construction and the pointers never slip. Once the FSM has locked, the fill level is frozen for the entire run and the buffer contributes a constant latency. The only effect able to move the operating point is a slow phase drift of the analog path, dominated by temperature, which can displace the lock point by one or two PI steps (a few tens of picoseconds) over hours. The firmware monitors \texttt{txbufstatus[0]} continuously and reports a flag toggle as a link-quality event; re-running the alignment FSM is only necessary after a reset, and is performed automatically at every link bring-up.

\section{The Front-End Board}
\label{sec:slave}
The front-end board is both the destination and a repeater of the synchronization stream: it recovers clock and data from the upstream fiber, uses them locally, and retransmits them to the next board of the chain. Its timing problem is therefore different from the concentrator one, and so is its solution.

\subsection{Datapath and Clock Recovery}
\label{sec:fersdatapath}
The FERS slave datapath is shown in Fig.~\ref{fig:fersdatapath}. The incoming optical bitstream is processed by a Xilinx Artix-7 GTP transceiver that recovers both the data and the embedded clock through its clock-and-data-recovery (CDR) block. The recovered 156.25~MHz clock inherits the data-dependent jitter introduced by the CDR; this jitter would accumulate hop-by-hop if the recovered clock were used directly as transmit reference for the next stage. To break this accumulation, the recovered clock is exported to an external zero-delay jitter-cleaner PLL built around a low-jitter voltage-controlled crystal oscillator (VCXO). The PLL output is distributed simultaneously to the GTP transmitter reference and to the FPGA fabric, including the TDC and ADC blocks. As a result, the same cleaned 156.25~MHz drives both the local logic and the ASICs and the downstream transmitter, locking the transmit frequency to the receive frequency and enforcing a well-defined retransmission relationship at every node along the daisy chain.

\begin{figure*}[!t]
\centering
\resizebox{7.0in}{!}{%
\begin{tikzpicture}
  \node[extd] (sfpin) at (0,0) {SFP\\RX};
  \node[mgt, minimum width=20mm] (rx) at (3.0,0) {GTP RX\\\footnotesize CDR + SIPO};
  \node[logic, minimum width=20mm] (dec) at (6.4,0) {comma align\\8b/10b decode};
  \node[logic, minimum width=20mm] (prot) at (9.8,0) {protocol FSM\\\footnotesize train insert};
  \node[logic, minimum width=20mm] (t0) at (13.2,0) {$T_0$ detect\\+ regeneration};
  \node[mgt, minimum width=20mm] (tx) at (16.6,0) {GTP TX\\\footnotesize PISO + buffer};
  \node[extd] (sfpout) at (19.6,0) {SFP\\TX};
  \foreach \a/\b in {sfpin/rx, rx/dec, dec/prot, prot/t0, t0/tx, tx/sfpout} {\draw[ar] (\a) -- (\b);}
  \node[extd, minimum width=30mm, minimum height=12mm] (jc) at (8.0,-2.8)
       {\textbf{external zero-delay PLL}\\jitter cleaner (VCXO)};
  \draw[clkar] (rx.south) |- node[pos=0.72,below,lbs]{recovered 156.25 MHz (jittered)} (jc.west);
  \draw[clkar] (jc.east) -| node[pos=0.28,below,lb]{cleaned 156.25 MHz} (tx.south);
  \node[logic, minimum width=26mm] (fabric) at (14.6,-4.6) {FPGA fabric\\PicoTDC, ADC, counters};
  \draw[clkar] (jc.south) |- (fabric.west);
\end{tikzpicture}}
\caption{Block diagram of the FERS slave datapath. The recovered clock is cleaned by an external zero-delay PLL and reused as the transmit reference, ensuring phase-coherent retransmission down the ring. The $T_0$ control word bypasses the packet layer and is regenerated at fixed latency by the mechanism of Section~\ref{sec:t0regen}.}
\label{fig:fersdatapath}
\end{figure*}

The clock recovery chain inside the FPGA is detailed in Fig.~\ref{fig:fersrecovery}. The GTP transceiver outputs 20-bit raw data words; the receiver searches the dual K28.5 comma pattern across two consecutive words to obtain a 40-bit alignment, which suppresses the 180$^{\circ}$ phase ambiguity inherent to the symmetric 10-bit comma. The transmitter forces alternating running disparity (RD+, RD$-$) on the two K28.5 symbols to make the dual-comma pattern unique on the line. A keep-alive mechanism re-issues the dual comma every 9.6~ms; if the receiver does not see commas within this window, the link is declared lost and the CDR is reset.

A further constraint is enforced on the word-alignment procedure, and it is essential to the reproducibility of the whole chain. Alignment is performed by issuing \texttt{RXSLIDE} pulses in physical-medium-attachment (PMA) mode, each of which delays the data by one unit interval (UI, 320~ps); the recovered parallel clock, however, is produced by a half-rate double-data-rate divider and moves only every \emph{second} slide. An odd total number of slides therefore corresponds to the opposite sampling edge: there are two families of perfectly comma-aligned lock states, separated by exactly 1~UI in the phase of the recovered clock with respect to the serial stream, and the family reached after a reset is random, because the parallelization boundary falls at a random position modulo the 20-bit word. Since the recovered clock is the time base of the entire downstream chain (jitter cleaner, transmit reference, fabric, TDC), a lock in the ``odd'' family would displace all subsequent timing by 1~UI. The receiver FSM therefore counts the slides issued since reset and rejects any lock reached with an odd count, re-issuing a receiver soft reset and repeating the hunt; the procedure converges in about two attempts on average and guarantees that every board, at every power-up, locks in the same phase family.

\begin{figure}[!t]
\centering
\resizebox{3.45in}{!}{%
\begin{tikzpicture}
  \node[mgt, minimum width=32mm] (cdr) at (0,0) {GTP CDR / PMA\\\footnotesize 20-bit raw output};
  \node[logic, minimum width=38mm] (align) at (0,-1.7) {dual-comma search (\texttt{RXSLIDE})\\\footnotesize \textbf{even slide count only}};
  \node[logic, minimum width=32mm] (dec) at (0,-3.4) {8b/10b decode in fabric};
  \node[logic, minimum width=36mm] (ka) at (0,-5.1) {keep-alive: comma every 9.6 ms};
  \foreach \a/\b in {cdr/align, align/dec, dec/ka} {\draw[ar] (\a) -- (\b);}
  \draw[ar] (ka.east) -- (3.3,-5.1) -- node[right,lb,align=left]
       {comma timeout:\\CDR reset} (3.3,0) -- (cdr.east);
  \draw[ar] (align.west) -- (-3.3,-1.7) -- node[left,lb,align=right]
       {odd count:\\soft reset\\and retry} (-3.3,0) -- (cdr.west);
  \node[extd, minimum width=40mm] (jc) at (0,-7.1)
       {external jitter cleaner (VCXO)\\\footnotesize zero-delay PLL};
  \node[logic, minimum width=40mm] (dst) at (0,-8.9)
       {TX reference, fabric, PicoTDC, ADC};
  \draw[clkar] (cdr.south west) -- (-5.4,-0.45) -- node[left,lb,align=right]
       {recovered\\clock} (-5.4,-7.1) -- (jc.west);
  \draw[clkar] (jc) -- node[right,lb]{cleaned 156.25 MHz} (dst);
\end{tikzpicture}}
\caption{FERS slave clock recovery. The transceiver CDR delivers a jittered 156.25~MHz clock which is filtered by an external jitter-cleaner PLL and distributed to the ASICs, the fabric and the transmit serializer. Word alignment accepts only locks reached with an even number of \texttt{RXSLIDE} pulses, which removes the 1~UI ambiguity of the half-rate deserializer and makes the recovered-clock phase reproducible at every power-up.}
\label{fig:fersrecovery}
\end{figure}

The alignment loop of Section~\ref{sec:master} makes the concentrator transmitters deterministic, but it does not, by itself, make the whole chain deterministic. The reason is a structural asymmetry between master and slave that is worth stating explicitly, because it dictates two different solutions.

\subsection{Why the Master Needs No TDC and the Slave Does}
On the concentrator, $T_0$ is \emph{born synchronous}: it is generated and sampled in the very clock domain that feeds the serializers, and it never crosses a clock-domain boundary on its way to the fiber. The only unknowns are the phases of \emph{periodic clocks} --- the two QPLL outputs --- and the phase of a periodic clock is exactly what the half-full flag measures locally, and what the DDMTD of Section~\ref{sec:ddmtd} measures across concentrators. Moreover, the randomness of a locked PLL at power-up is confined to its output divider chain: the QPLL voltage-controlled oscillator is phase-locked to the shared reference, so in steady state its edges sit at a fixed, picosecond-class offset from the reference, and what starts at random is only which oscillator edge the dividers begin counting from. Since a single reset FSM releases the datapaths, dividers and buffers of all eight lanes together, and since both oscillators are locked to the same reference, the dividers of the two quads start in the same state and the buffers come up with the same fill level. This is why buffer-plus-PI is sufficient on the master, and why the measured inter-lane residual stays within one PI step.

On a front-end board the situation is different. The $T_0$ arrives as an \emph{event} --- a control word decoded in the recovered-clock domain --- and must be re-emitted in the transmit domain. Even though the two domains are frequency-locked through the jitter cleaner, they are not phase-identical, and the pulse must therefore traverse a CDC synchronizer. A standard multi-flip-flop pulse synchronizer quantizes with a latency of three, four or five word clocks depending on the sampled phase, i.e.\ it introduces an ambiguity of up to two word periods (12.8~ns) that varies from board to board and, critically, from one power-up to the next. This is an \emph{event}-domain problem, not a periodic-clock problem, and neither the half-full flag nor the DDMTD can observe it.

\subsection{Multiphase Oversampling TDC and Fixed-Latency Scheduler}
\label{sec:t0regen}
The solution adopted on the front-end boards is to time-stamp the incoming $T_0$ edge directly on the transmit time base and to re-emit the pulse at a fixed, programmable distance from that time stamp, so that the CDC synchronizer is removed from the signal path altogether (Fig.~\ref{fig:slavetdc}).

The time stamp is produced by a multiphase oversampling TDC implemented in the fabric. A PLL multiplies the 156.25~MHz transmit clock to two 312.5~MHz clocks in quadrature ($0^{\circ}$ and $90^{\circ}$); four sampling chains capture the input on the rising \emph{and} falling edges of both phases, giving $2\times2\times2 = 8$ uniformly spaced sampling instants per word period, i.e.\ a bin width of
\begin{equation}
\label{eq:tdcbin}
\Delta t = \frac{T_W}{8} = \frac{1}{8 f_W} = 0.8~\mathrm{ns}.
\end{equation}
This architecture was preferred over a delay-line (tapped carry chain) TDC because the required single-shot resolution is modest, while the bins of a multiphase sampler are uniform \emph{by construction} --- they derive from PLL phase relationships rather than from logic propagation delays --- so the converter requires no per-device calibration and behaves identically on every board programmed with the same bitstream. Each chain starts with a two-flip-flop synchronizer; the falling-edge chains are re-registered on the rising edge of their own clock so that every transfer toward the transmit domain has at least 2.4~ns of analyzed margin. All the clocks involved are PLL-correlated, so these are timed paths, not clock-domain crossings.

The eight samples are recomposed each cycle into an 8-bit window spanning exactly one word period, and the time stamp of an edge is
\begin{equation}
\label{eq:tdcstamp}
t = 8\,n_{\rm coarse} + \mathrm{first\_one}(\mathrm{win}), \qquad \mathrm{LSB} = 0.8~\mathrm{ns},
\end{equation}
where $n_{\rm coarse}$ is a free-running word-clock counter. Two \emph{identical} sampler instances are used, which makes the measurement differential: channel~A digitizes the $T_0$ decoded in the recovered-clock domain, channel~B the output of the legacy CDC synchronizer. Every offset common to the two chains --- synchronizer depth, window recomposition, PLL phase --- cancels exactly in the difference $\Delta_{\rm CDC} = t_B - t_A$, which therefore measures the true synchronizer latency and is retained as a permanent on-line diagnostic.

The regeneration itself does not use channel~B at all. When the TDC detects the rising edge of channel~A, a scheduler is armed at $n_{\rm coarse} + N_{\rm target}$, with $N_{\rm target}$ a programmable number of word clocks (default~8); when the free-running counter reaches that value, the $T_0$ pulse is emitted. The emitted pulse is therefore always at the same distance from the receive-domain edge, whatever the CDC synchronizer did on that particular event: the variability is absorbed on the \emph{same} $T_0$, not corrected on the next one, because the decision reference is the received edge itself. The data path is delayed by a fixed tap of $N_{\rm target}-2$ word clocks so that the injected $T_0$ control word replaces the same stream word it would have replaced on a direct path, with no discontinuity in the data flow. The total latency of the regeneration path,
\begin{equation}
\label{eq:t0lat}
L_{T_0} = \left( N_{\rm target} + L_{\rm det} \right) T_W \approx \left( N_{\rm target} + 2.5 \right) T_W,
\end{equation}
with $L_{\rm det}$ the constant edge-detection pipeline, is identical on all boards and at every power-up.

\begin{figure*}[!t]
\centering
\resizebox{7.0in}{!}{%
\begin{tikzpicture}
  \node[logic, minimum width=22mm] (det) at (0,0) {$T_0$ decode\\\footnotesize recovered-clock domain};
  \node[hl, minimum width=22mm] (smpA) at (4.2,0) {\textbf{sampler A}\\\footnotesize 8 bins $\times$ 0.8 ns};
  \node[hl, minimum width=24mm] (sch) at (8.6,0) {\textbf{scheduler}\\\footnotesize emit at $n+N_{\rm target}$};
  \node[logic, minimum width=22mm] (inj) at (13.0,0) {$T_0$ word\\injection};
  \node[mgt, minimum width=20mm] (tx) at (17.0,0) {GTP TX\\\footnotesize buffer + PI};
  \foreach \a/\b in {det/smpA, smpA/sch, sch/inj, inj/tx} {\draw[ar] (\a) -- (\b);}
  \node[lbs, above=1mm of smpA] {edge time stamp $t_A$};
  \node[logic, minimum width=24mm] (xpm) at (4.2,-2.6) {CDC synchronizer\\\footnotesize 3/4/5 word clocks};
  \node[logic, minimum width=22mm] (smpB) at (8.6,-2.6) {sampler B\\\footnotesize identical to A};
  \node[logic, minimum width=24mm] (mon) at (13.0,-2.6) {$\Delta_{\rm CDC} = t_B - t_A$\\\footnotesize monitor only};
  \draw[ar] (det.south) |- (xpm.west);
  \draw[ar] (xpm) -- (smpB);
  \draw[ar] (smpB) -- node[above,lb]{$t_B$} (mon);
  \draw[ar] (smpA.south) -- node[pos=0.62, above left=-0.5mm, lb]{$t_A$} (mon.north west);
  \node[logic, minimum width=24mm] (dly) at (13.0,2.3) {data delay\\\footnotesize fixed $N_{\rm target}-2$ taps};
  \draw[ar] (dly) -- (inj);
  \node[lb, anchor=north west, text width=96mm] at (-1.4,-4.2)
       {The CDC synchronizer is digitized and monitored but never drives the
        link: the emitted $T_0$ is referenced to the received edge, so its
        latency is fixed by construction.};
\end{tikzpicture}}
\caption{Deterministic $T_0$ regeneration on a FERS board. The multiphase TDC time-stamps the received $T_0$ edge on the transmit time base with 0.8~ns bins; a scheduler re-emits the pulse exactly $N_{\rm target}$ word clocks later, and the data path is delayed by a matching fixed tap. The conventional CDC synchronizer, whose latency varies between three and five word clocks, is digitized by an identical second sampler and retained only as an on-line diagnostic.}
\label{fig:slavetdc}
\end{figure*}

\section{Multi-Concentrator Synchronization}
\label{sec:multicnc}
A single concentrator instruments up to 128 boards. Larger systems are built by cascading concentrators, which raises the same phase problem one level up: the transmit clocks of two concentrators are derived from two different AD9545 devices and, even when both are locked to the same reference, their outputs are not phase-identical. Here, unlike the intra-concentrator case, the quantity to be steered \emph{is} the input reference of the device, and the AD9545 programmable delay provides exactly that degree of freedom. What is needed is a way to measure the residual offset with picosecond resolution, which is the role of the DDMTD.

The hierarchy is established as shown in Fig.~\ref{fig:multicnc}: the master concentrator distributes its 156.25~MHz reference to one or more slave concentrators through the synchronization chain input, and each slave locks its input PLL to that reference and drives its own two QPLL groups exactly as the master does. Within every concentrator of the hierarchy the lane alignment loop of Section~\ref{sec:master} runs unchanged, so the only term left to correct is the offset between the references of different crates.

\begin{figure}[!t]
\centering
\resizebox{3.45in}{!}{%
\begin{tikzpicture}
  \node[extd, minimum width=26mm, minimum height=12mm] (m) at (0,0)
       {\textbf{MASTER} DT5215\\\footnotesize local TCXO / GPS};
  \node[logic, minimum width=15mm] (mq0) at (3.6,0.6) {QUAD 0};
  \node[logic, minimum width=15mm] (mq1) at (3.6,-0.6) {QUAD 1};
  \node[blk, minimum width=22mm] (mr0) at (6.9,0.6) {4 rings};
  \node[blk, minimum width=22mm] (mr1) at (6.9,-0.6) {4 rings};
  \draw[ar] (m.east|-mq0) -- (mq0);  \draw[ar] (m.east|-mq1) -- (mq1);
  \draw[ar] (mq0) -- (mr0);  \draw[ar] (mq1) -- (mr1);
  \node[extd, minimum width=26mm, minimum height=12mm] (s) at (0,-2.9)
       {\textbf{SLAVE} DT5215\\\footnotesize locked to master};
  \node[logic, minimum width=15mm] (sq0) at (3.6,-2.3) {QUAD 0};
  \node[logic, minimum width=15mm] (sq1) at (3.6,-3.5) {QUAD 1};
  \node[blk, minimum width=22mm] (sr0) at (6.9,-2.3) {4 rings};
  \node[blk, minimum width=22mm] (sr1) at (6.9,-3.5) {4 rings};
  \draw[ar] (s.east|-sq0) -- (sq0);  \draw[ar] (s.east|-sq1) -- (sq1);
  \draw[ar] (sq0) -- (sr0);  \draw[ar] (sq1) -- (sr1);
  \draw[clkar] (m.south) -- node[right,lbs]{156.25 MHz ref.} (s.north);
  \node[logic, minimum width=30mm] (dd) at (0,-5.2) {DDMTD: slave vs master phase};
  \draw[ar] ($(s.south)+(-0.5,0)$) -- ($(dd.north)+(-0.5,0)$);
  \draw[ar] ($(dd.north)+(0.5,0)$) -- node[right,lbs]{AD9545 delay correction}
       ($(s.south)+(0.5,0)$);
  \node[lbs, anchor=north west, text width=58mm] at (-2.2,-6.1)
       {Each ring hosts up to 16 FERS boards; 8 rings per concentrator.};
\end{tikzpicture}}
\caption{Multi-concentrator topology. A master concentrator distributes its 156.25~MHz reference to one or more slave concentrators. Each concentrator drives two QPLL groups, each feeding four TD-Link rings. The DDMTD measurement closes a correction loop on the programmable delay of the slave AD9545, aligning the slave reference to the master.}
\label{fig:multicnc}
\end{figure}

\subsection{The DDMTD Phase Detector}
\label{sec:ddmtd}
The Digital Dual Mixer Time Difference (DDMTD) is a fully-digital phase detector, originally adopted in the White Rabbit project \cite{b5,b3}, that achieves sub-picosecond resolution using only standard FPGA logic. Its principle is illustrated in Fig.~\ref{fig:ddmtd}. The two clocks to be compared, ${\rm clk}_{\rm A}$ and ${\rm clk}_{\rm B}$, both at $f_{\rm in}$, are sampled by D flip-flops clocked by a third clock ${\rm clk}_{\rm DDMTD}$ at frequency $f_{\rm in}+\Delta f$, with $\Delta f \ll f_{\rm in}$. Each flip-flop output is a low-frequency square wave (the beat) at frequency $\Delta f$. The rising edges of the two beat signals are separated in time by
\begin{equation}
\label{eq:ddmtd}
\Delta T_{\rm beat} = \Delta\Phi \cdot \frac{f_{\rm in}}{\Delta f} ,
\end{equation}
where $\Delta\Phi$ is the phase difference between the two input clocks expressed as a time interval. The native phase resolution is therefore magnified by a factor $f_{\rm in}/\Delta f$, and the beat-edge separation can be measured with an interval counter clocked at the FPGA fabric frequency. With $f_{\rm in}=156.25$~MHz and $\Delta f=15.625$~kHz, the magnification factor is $10^4$, giving a theoretical resolution of about 0.6~ps. Metastability glitches at the beat transitions are removed by a digital deglitcher upstream of the interval counter.

\begin{figure}[!t]
\centering
\resizebox{3.45in}{!}{%
\begin{tikzpicture}
  \node[logic, minimum width=21mm] (ffa) at (0,1.5) {D flip-flop A};
  \node[logic, minimum width=21mm] (ffb) at (0,-1.5) {D flip-flop B};
  \draw[ar] (0,3.1) -- node[right,lb]{${\rm clk}_{\rm A}$ ($f_{\rm in}$)} (ffa.north);
  \draw[ar] (0,-3.1) -- node[right,lb]{${\rm clk}_{\rm B}$ ($f_{\rm in}$)} (ffb.south);
  \node[extd, minimum width=24mm, align=center] (hp) at (-3.7,0)
       {helper PLL\\${\rm clk}_{\rm DDMTD}$\\$= f_{\rm in} + \Delta f$};
  \draw[very thick] (hp.east) -- (-1.9,0);
  \draw[clkar] (-1.9,0) |- (ffa.west);
  \draw[clkar] (-1.9,0) |- (ffb.west);
  \node[logic, minimum width=21mm] (dg) at (3.3,0) {deglitcher};
  \draw[ar] (ffa.east) -| node[pos=0.3,above,lb]{beat A ($\Delta f$)} (dg.north);
  \draw[ar] (ffb.east) -| node[pos=0.3,below,lb]{beat B ($\Delta f$)} (dg.south);
  \node[logic, minimum width=25mm] (cnt) at (3.3,-2.9) {interval counter};
  \draw[ar] (dg.east) -- (5.1,0) -- (5.1,-2.9) -- (cnt.east);
  \node[lb, anchor=north west, text width=58mm] at (-4.9,-4.1)
       {$\Delta T_{\rm beat} = \Delta\Phi \cdot f_{\rm in}/\Delta f$:
        magnification $10^4$, resolution $\approx 0.6$ ps};
\end{tikzpicture}}
\caption{Principle of the DDMTD phase detector. The two clocks to be compared are sampled by D flip-flops clocked at $f_{\rm in}+\Delta f$. The phase information is time-magnified by a factor $f_{\rm in}/\Delta f$ and recovered with a digital counter, achieving sub-picosecond resolution with standard FPGA logic only.}
\label{fig:ddmtd}
\end{figure}

In TD-Link, the DDMTD is used in two scenarios. The first, internal to a concentrator, measures the relative phase between the transmit word clock of each quad and the fabric reference clock; this measurement verifies the lane alignment loop and logs link-quality metrics. The second, external, compares the recovered clock of a slave concentrator against the reference distributed by the master, and drives a closed-loop correction applied to the programmable delay of the AD9545 input PLL \cite{b12}, operated in zero-delay mode. The closed-loop bandwidth is set well below the residual phase noise of the recovered clock, and the loop converges within a few seconds after link lock.

\section{Per-Hop Latency Budget}
\label{sec:budget}
The mechanisms described above are not independent fixes: each addresses the timing uncertainty at a different scale and hands the residue to the next. Table~\ref{tab:budget} summarizes the decomposition of the per-hop latency and the mechanism that pins each term. The net result is that every board of a chain inserts a delay that is large but exactly known, and reproducible to a few tens of picoseconds at every power-up; the $T_0$ arrival time at board $n$ of a ring is the analytically known sum $t_0 + n L_{\rm hop} + \sum \tau_{\rm fiber}$. Each mechanism is also observable: the CDC monitor, the PI phase readback, the odd-slide counter and the jitter-cleaner lock detector are all exported to the Janus framework \cite{b11}, so a hop that degraded its calibration is detected rather than silently absorbed.

\begin{table}[!t]
\caption{Per-hop latency terms and stabilizing mechanisms}
\label{tab:budget}
\centering
\setlength{\tabcolsep}{4pt}
\begin{tabular}{@{}L{54pt}L{54pt}L{66pt}L{40pt}@{}}
\toprule
Term & Raw uncertainty & Mechanism & Residual \\
\midrule
$T_0$ domain crossing & 3--5 word clocks (12.8 ns) & multiphase TDC $+$ fixed-latency scheduler & exact integer \\
\addlinespace[4pt]
TX elastic buffer & random fill at reset & half-full flag $+$ PI sweep (FSM) & $\approx$ 1 PI step \\
\addlinespace[4pt]
Word alignment & 1 UI (320 ps) & even-slide-only \texttt{RXSLIDE} policy & 0 \\
\addlinespace[4pt]
Clock phase & accumulating CDR jitter & external zero-delay jitter cleaner & $\approx$ 20 ps \\
\bottomrule
\end{tabular}
\end{table}

\section{Experimental Results}
\label{sec:results}
The synchronization performance of TD-Link has been characterized using two DT5215 Data Concentrators and several CAEN~A5203 FERS-5200 boards equipped with the CERN PicoTDC \cite{b4}. The PicoTDC provides a single-channel time-to-digital resolution below 10~ps, which makes it well-suited to measure the timing jitter of the optical link. Pairs of correlated pulses were injected simultaneously into two channels on different A5203 boards, and the distribution of the time difference between the two time stamps was acquired for each configuration. The standard deviation $\sigma$ of this distribution is the figure of merit reported below; it includes the intrinsic resolution of the two PicoTDC channels, the residual phase noise of the recovered clock, and the contribution of the alignment loops. Every configuration was repeated across ten power cycles.

\subsection{Baseline: Coaxial Clock Reference}
\label{sec:resbaseline}
\begin{figure*}[!t]
\centerline{%
\includegraphics[height=2.15in]{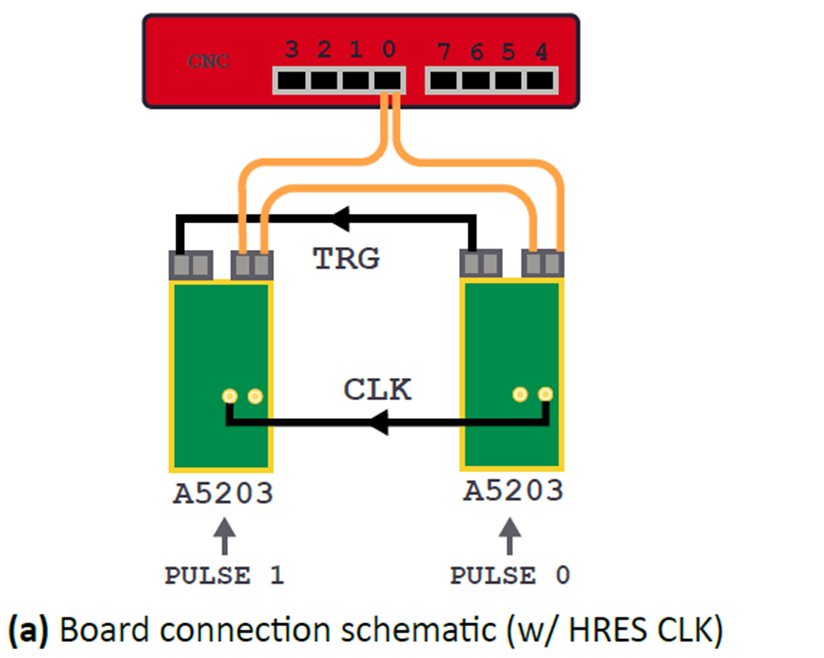}\hfill
\includegraphics[height=2.15in]{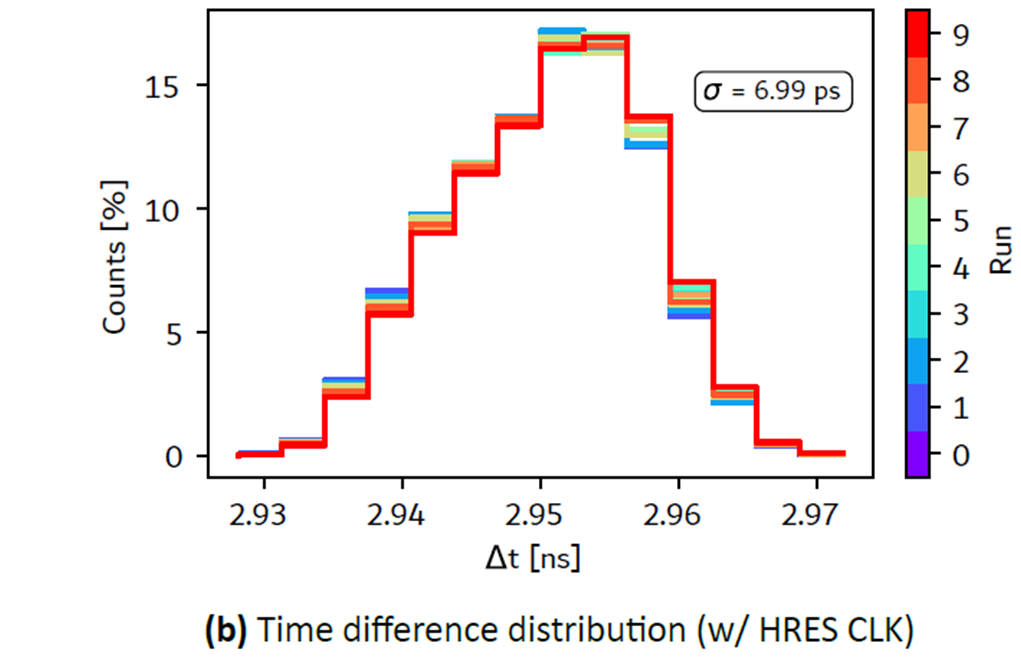}}
\caption{Baseline jitter measurement with the timing reference distributed via a coaxial cable. (a)~Test setup. (b)~Time-difference distribution overlaid for ten power cycles. The measured $\sigma \simeq 7$~ps represents the intrinsic jitter floor of the FERS front-end, dominated by TDC quantization noise.}
\label{fig:rescable}
\end{figure*}

As a baseline, two FERS boards were connected through a short coaxial cable carrying a common timing reference (Fig.~\ref{fig:rescable}). This configuration removes any optical clock-distribution contribution and exposes the intrinsic jitter floor of the FERS front-end. The measured time-difference distribution has $\sigma \simeq 7$~ps, dominated by the quantization noise of the two TDC channels in quadrature, and is stable across all ten power cycles (the overlaid curves in Fig.~\ref{fig:rescable}b). A reference measurement with both channels on the \emph{same} board gives $\sigma \simeq 5.3$~ps, which is the single-board floor.

\subsection{Optical Configurations}
\label{sec:resoptical}
Three TD-Link configurations of increasing complexity were then characterized, exercising progressively more of the synchronization chain.

\begin{figure*}[!t]
\centerline{%
\includegraphics[height=2.15in]{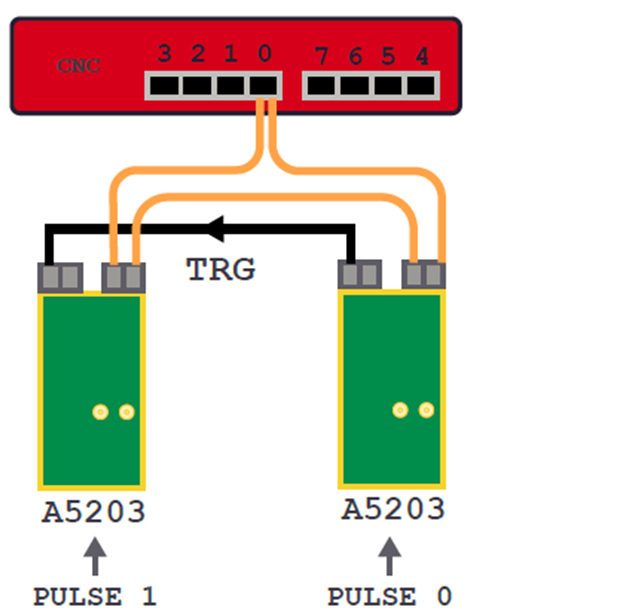}\hfill
\includegraphics[height=2.15in]{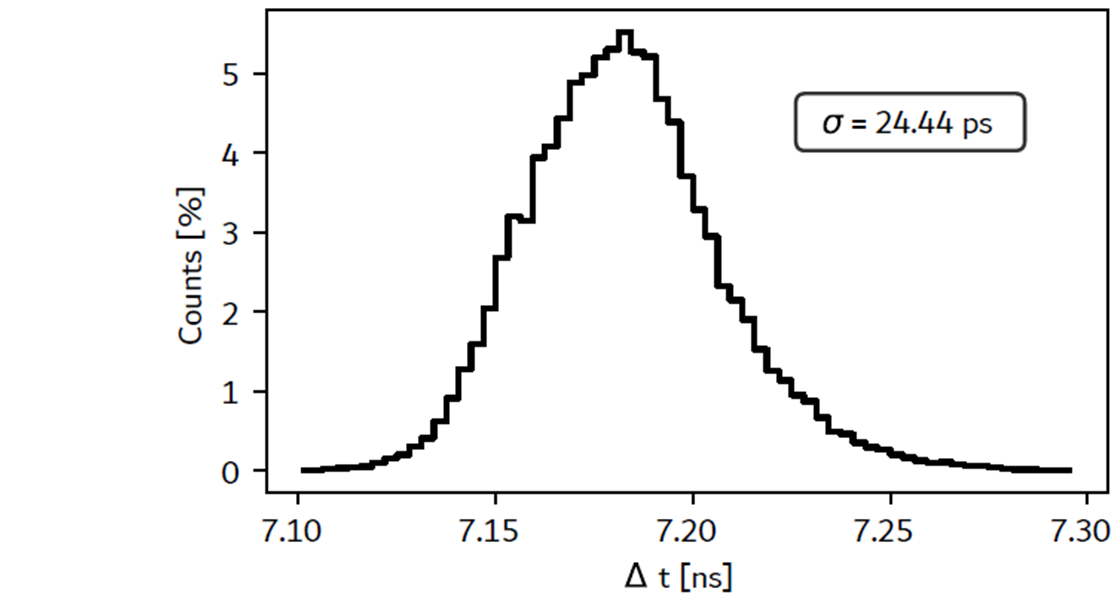}}
\caption{Time-difference distribution for two FERS boards on the same TD-Link ring (hops 2 and 3), with the clock distributed via the optical daisy chain. The measured $\sigma \simeq 24$~ps is dominated by the per-board clock-recovery and jitter-cleaning chain.}
\label{fig:reschain}
\end{figure*}

\begin{figure*}[!t]
\centerline{%
\includegraphics[height=2.15in]{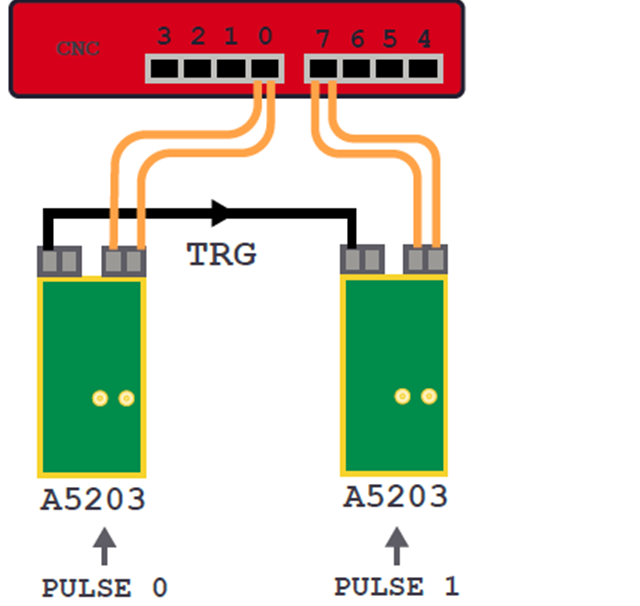}\hfill
\includegraphics[height=2.15in]{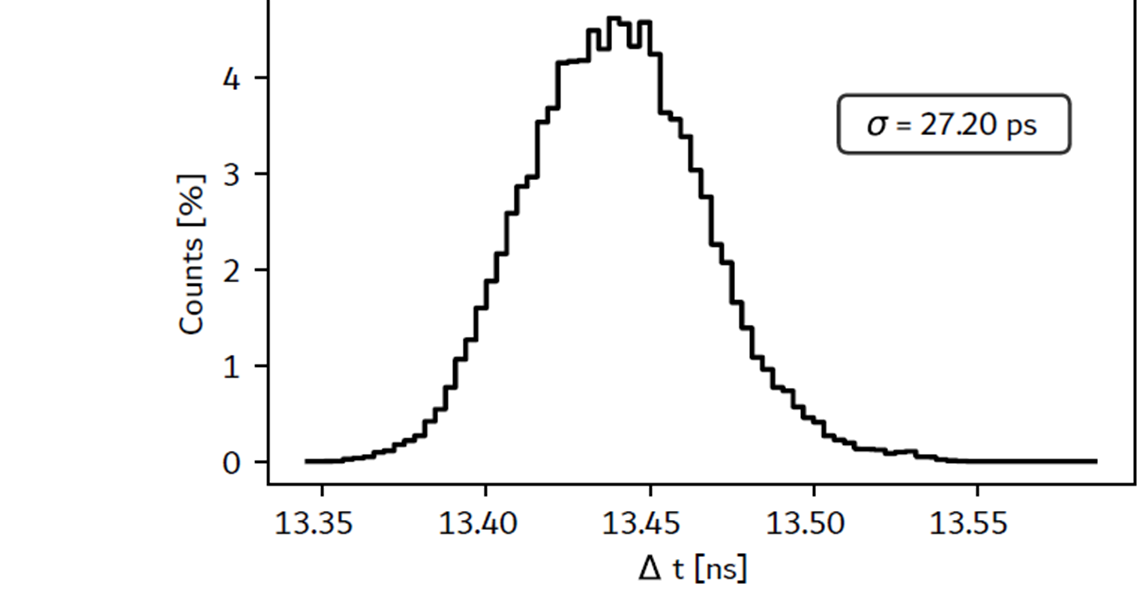}}
\caption{Time-difference distribution for two FERS boards on different TD-Link rings of the same concentrator (lanes 0 and 7, two different QPLL quads). The measured $\sigma \simeq 27$~ps is comparable to the same-ring case, confirming the effectiveness of the deterministic TX buffer alignment loop.}
\label{fig:resquads}
\end{figure*}

\begin{figure*}[!t]
\centerline{%
\includegraphics[height=2.15in]{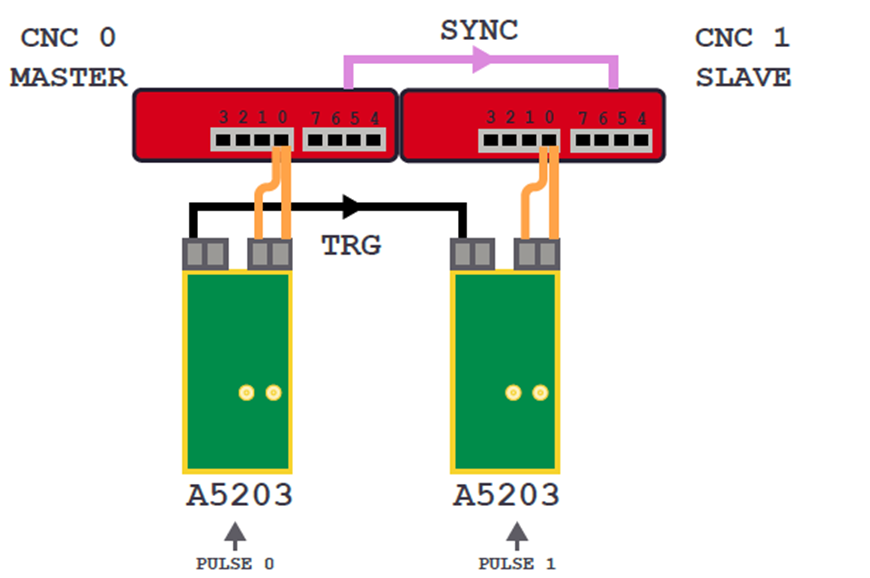}\hfill
\includegraphics[height=2.15in]{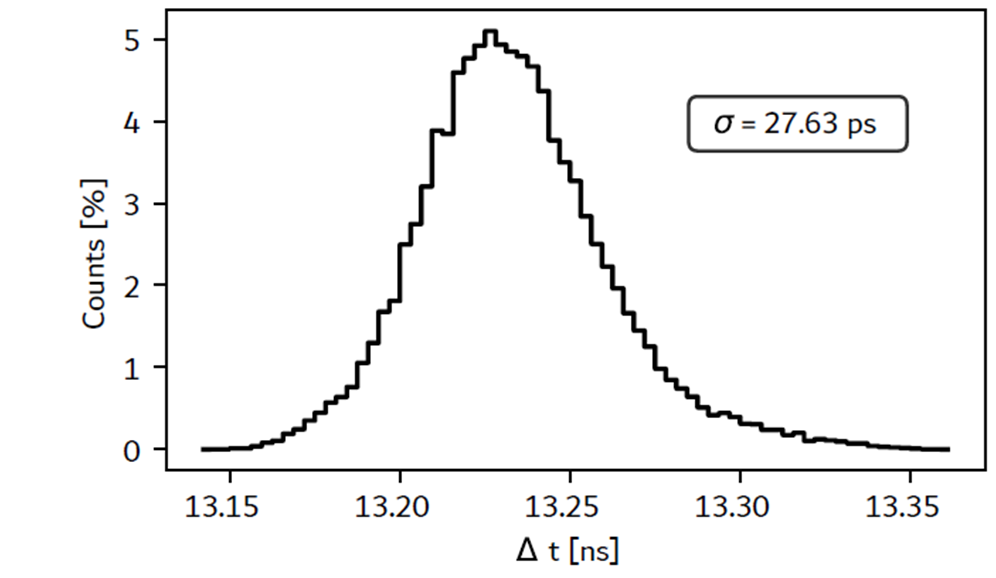}}
\caption{Time-difference distribution for two FERS boards belonging to different concentrators (master and slave). The full synchronization chain is exercised, including the inter-concentrator DDMTD correction. The measured $\sigma \simeq 28$~ps is only marginally larger than the single-concentrator case.}
\label{fig:res2cnc}
\end{figure*}

In the \emph{same-ring} configuration, two FERS boards belong to the same daisy chain --- one at the second hop, the other at the third (Fig.~\ref{fig:reschain}) --- so both receive the timing clock through the optical chain and the measurement probes the per-hop clock recovery and jitter cleaning. The measured $\sigma \simeq 24$~ps. In the \emph{same-quad, different-ring} configuration, the two boards sit on two rings served by the same QPLL, and $\sigma \simeq 25.4$~ps. In the \emph{different-quad} configuration (Fig.~\ref{fig:resquads}) the two rings are served by two different QPLLs of the same concentrator (lanes~0 and~7), so the deterministic TX buffer alignment loop of Section~\ref{sec:master} is the only mechanism bringing the two recovered clocks into phase; the measured $\sigma \simeq 27$~ps. Finally, in the \emph{two-concentrator} configuration (Fig.~\ref{fig:res2cnc}) the boards are attached to a master and a slave concentrator, exercising the full chain including the DDMTD-based correction on the slave AD9545; the measured $\sigma \simeq 28$~ps.

\subsection{Jitter Budget}
\label{sec:resbudget}
Table~\ref{tab:jitter} collects the measured $\sigma$ for every configuration, together with the residual obtained by subtracting the 7~ps front-end floor in quadrature. The picture is unambiguous: the step from the coaxial reference to the optical chain costs about 23~ps in quadrature, while every subsequent step --- adding a quad crossing, then a concentrator crossing --- costs only a few picoseconds more. The dominant contribution to the TD-Link figure is therefore the phase noise of the per-board recovered-clock and jitter-cleaner chain, \emph{not} the TDC quantization (which dominates only the coaxial baseline) and not the alignment loops, whose residual is buried under it. Repeating each configuration with the coaxial reference restores $\sigma \simeq 7$~ps in all topologies, confirming that the additional jitter enters with the optical clock distribution and not with the front-end or the readout path.

\begin{table}[!t]
\caption{Measured synchronization jitter by configuration}
\label{tab:jitter}
\centering
\setlength{\tabcolsep}{4pt}
\begin{tabular}{@{}L{94pt}L{34pt}r@{\hskip 10pt}r@{}}
\toprule
Configuration & Clock source & $\sigma$ (ps) & Residual$^{\rm a}$ (ps) \\
\midrule
Two channels, same board      & coaxial & 5.3 & --- \\
\addlinespace[2pt]
Two boards                    & coaxial & 7.0 & --- \\
\addlinespace[2pt]
Same ring (hops 2--3)         & TD-Link & 24  & 23 \\
\addlinespace[2pt]
Same quad, different rings    & TD-Link & 25.4 & 24 \\
\addlinespace[2pt]
Different quads, same concentrator & TD-Link & 27 & 26 \\
\addlinespace[2pt]
Two concentrators (master/slave) & TD-Link & 28 & 27 \\
\addlinespace[2pt]
Any of the above              & coaxial & $\simeq 7$ & --- \\
\bottomrule
\addlinespace[2pt]
\multicolumn{4}{@{}L{210pt}@{}}{\footnotesize $^{\rm a}$ After quadrature subtraction of the 7~ps two-board front-end floor.}
\end{tabular}
\end{table}

\subsection{Reproducibility Across Power Cycles}
\label{sec:respower}
Reproducibility was evaluated at two levels. First, the width of the distribution: repeating each measurement across ten power cycles gives a cycle-to-cycle spread of the $\sigma$ values better than 1~ps in all configurations, i.e.\ the \emph{precision} of the link is fully reproducible.

Second, and more demanding, the \emph{absolute board-to-board skew} recovered after each cold boot. This is the quantity that the deterministic chain must preserve, and it is where the mechanisms of Sections~\ref{sec:master} and~\ref{sec:t0regen} are actually tested. It is also a quantity for which the PicoTDC measurement is not the appropriate instrument: a histogram of time differences requires a long acquisition for every boot, which makes a campaign of thousands of power cycles impractical, and its mean is affected by the per-channel offsets of the two TDC channels. The reproducibility campaign was therefore performed directly in the time domain. Each FERS board exports its regenerated $T_0$ on a front-panel monitor output; the outputs of two boards were connected to a fast real-time oscilloscope and the skew between the two edges was measured after every cold boot, repeating the procedure on two boards of the same ring and on two boards served by different rings and different quads.

With only the concentrator-side alignment in place, the skew was reproducible across quads but could jump by one word clock (6.4~ns) on a front-end board between power cycles, because of the three-to-five-cycle ambiguity of the $T_0$ domain crossing described in Section~\ref{sec:t0regen}. With the TDC-based $T_0$ regeneration enabled, this term is removed by construction. The complete chain was subsequently exercised with more than one thousand cold boots of the full system, without any re-calibration between them: no word-clock jump was ever observed, and the boot-to-boot spread of the measured skew remained \textbf{below 20~ps} in all configurations. This figure is consistent with the 18.6~ps quantization of the concentrator PI step, which sets the granularity of the transmit alignment, and it is quoted conservatively: the spread observed on the oscilloscope traces is smaller, but we prefer to bound it by the quantization of the mechanism that produces it rather than by the resolution of the instrument used to observe it.

The dominant systematic effect identified is temperature drift over time scales of hours, which can move the lock point of the transmit buffer loop by one or two PI steps (a few tens of picoseconds) and is automatically re-converged at every link bring-up. A systematic characterization of the drift as a function of temperature has not been performed and is left for future work.

\section{Comparison with White Rabbit}
\label{sec:wr}
Table~\ref{tab:wr} summarizes the main features of TD-Link in comparison with White Rabbit \cite{b3}, the reference open-source protocol for sub-nanosecond timing in high-energy-physics and accelerator facilities. The two systems address overlapping but distinct deployment scenarios, and the comparison must be read with care, because the two figures of merit are different in nature. White Rabbit is specified in terms of \emph{accuracy}: the absolute offset of each endpoint with respect to the grandmaster, typically below 1~ns and reaching about 100~ps after calibration of the fiber and endpoint asymmetries \cite{b3,b6}. The TD-Link figures reported in this paper are a \emph{precision}: the standard deviation of the board-to-board time difference.

TD-Link, in fact, \emph{cannot} measure the absolute propagation delay to an individual node, and this is a structural consequence of the ring topology rather than an implementation choice. The stream returns to the concentrator only after traversing the entire chain, so the round-trip measurement performed at enumeration time yields the total latency of the ring; it carries no information about how that latency is distributed among the individual hops. Resolving the delay of a single node would require a dedicated return path for that node, which is precisely what a point-to-point topology provides and what the daisy chain trades away in exchange for its cabling economy. In TD-Link the per-node arrival time is therefore obtained \emph{by construction} rather than by measurement: because the per-hop latency is deterministic and identical on every board (Section~\ref{sec:budget}), the $T_0$ arrival time at board $n$ is $t_0 + n L_{\rm hop} + \sum \tau_{\rm fiber}$, where the fiber contribution is computed from the installed cable lengths and applied as a per-board correction in software. The accuracy of that correction is limited by the knowledge of the cable lengths, not by the link itself. The two numbers are therefore complementary rather than directly comparable, and no claim of superiority over White Rabbit is made here.

White Rabbit is built around switched star or tree topologies and dedicated switches; it is optimized for very large facilities where many independent endpoints share an absolute time reference over kilometers of single-mode fiber. TD-Link is optimized for the opposite scenario: up to a few thousand front-end boards physically chained within a compact region, typically the front-end volume of a single detector, where minimizing cabling and the number of optical interfaces is the leading design constraint. The daisy-chain topology is the source of both its main advantage --- timing distribution integrated in the readout hardware, no dedicated timing switch --- and its main limitation: a single board or fiber failure interrupts the ring, and the whole chain downstream of the fault is lost until the faulty element is bypassed or replaced. Deployments requiring fault tolerance at the single-node level are better served by a star topology. We also note that cascading was demonstrated in this work for two concentrators; the master/slave scheme is designed to extend further, but a quantitative characterization of deeper cascades remains to be performed.

\begin{table}[!t]
\caption{Comparison between TD-Link and White Rabbit}
\label{tab:wr}
\setlength{\tabcolsep}{3pt}
\begin{tabular}{|p{72pt}|p{78pt}|p{78pt}|}
\hline
Feature & TD-Link & White Rabbit \\
\hline
Figure of merit & precision (board-to-board $\sigma$) & accuracy (offset to grandmaster) \\
Reported value & 24--28 ps $\sigma$ & $<$1 ns, $\sim$100 ps calibrated \\
Absolute delay compensation & round-trip measured at enumeration, applied in software & continuous, in-protocol \\
Clock distribution & embedded in serial data & SyncE + PTP (IEEE 1588) \\
Topology & daisy-chain ring & star / tree (switches) \\
Link rate & 3.125 Gb/s per ring & 1 Gb/s (GbE) \\
Max nodes per link & 16 per ring, 8 rings per concentrator & $\sim$1000 (network) \\
Max distance & SFP 300 m per hop & up to 10 km SMF \\
Dedicated switch needed & no & yes (WR Switch) \\
Single-point failure & breaks the ring downstream & isolated to one endpoint \\
Target application & detector front-end DAQ & lab / accelerator timing \\
\hline
\end{tabular}
\end{table}

\section{Conclusion}
\label{sec:conclusion}
We have presented TD-Link, a custom 3.125~Gb/s optical communication protocol that combines high-throughput data readout and deterministic, sub-nanosecond timing synchronization on a single multidrop daisy-chain ring. The protocol carries detector data, control, configuration, and the global time reference $T_0$ within the same continuous serial stream, using a token-based train protocol for data and a dedicated fixed-latency path for $T_0$. Up to sixteen FERS-5200 boards can be chained on a single ring, with up to eight rings per Data Concentrator and multiple concentrators cascaded through a master/slave hierarchy.

The synchronization architecture rests on four pillars: a low-jitter master clock generated from a 10~MHz TCXO or from an external/GPS reference; a clean retransmission of the recovered clock at every FERS node via an external zero-delay PLL, combined with a word-alignment policy that rejects the ambiguous lock family and makes the recovered-clock phase reproducible; an FPGA-only alignment mechanism on the concentrator that exploits the half-full flag of the transmit elastic buffer as a one-bit phase detector driving the transmit PI, which resolves the dual-purpose conflict that prevents buffer bypass when lanes from different quads must be aligned; and, on each front-end board, a calibration-free multiphase oversampling TDC that time-stamps the incoming $T_0$ on the transmit time base so that a scheduler can regenerate it at fixed latency, eliminating the three-to-five-cycle ambiguity of the clock-domain crossing. A DDMTD circuit, implemented entirely in standard FPGA logic, provides the sub-picosecond phase measurement used for inter-concentrator alignment.

Experimental validation with PicoTDC-equipped FERS boards has demonstrated a board-to-board synchronization $\sigma$ of 24~ps on the same optical ring, 27~ps across transceiver quads and 28~ps across two cascaded concentrators, against an intrinsic front-end floor of 7~ps measured with a coaxial reference clock; the residual after quadrature subtraction shows that the dominant contribution is the per-board recovered-clock chain rather than the alignment loops. Reproducibility was assessed separately in the time domain, by comparing the $T_0$ monitor outputs of two boards on a fast oscilloscope after every power cycle: with the complete deterministic chain enabled, more than one thousand cold boots were executed without re-calibration, no word-clock jump was ever observed, and the boot-to-boot spread of the board-to-board skew stayed below 20~ps, the granularity of the phase-interpolator step. TD-Link therefore combines sub-30~ps timing precision with a scalable, low-infrastructure readout architecture, providing a practical alternative to switched timing networks for next-generation large-scale detector front-ends.

\end{document}